\documentclass[11pt]{article}

\usepackage[utf8]{inputenc}
\usepackage[T1]{fontenc}

\usepackage{amsmath,amssymb,amsthm}
\usepackage{bm}
\usepackage{mathrsfs}

\usepackage{graphicx}
\usepackage{float}
\usepackage{caption}
\usepackage{subcaption}

\usepackage{booktabs}
\usepackage{multirow}
\usepackage{array}

\usepackage{tabularx}
\usepackage{array}

\usepackage{hyperref}
\hypersetup{
    colorlinks=true,
    linkcolor=blue,
    citecolor=blue,
    urlcolor=blue
}

\usepackage[a4paper,margin=1in]{geometry}

\usepackage{setspace}
\theoremstyle{definition}

\theoremstyle{remark}

\title{
    \textbf{WIP: Bridging the Gap Between Instructional Design and Pedagogical Use: A Framework for Mathematics Educators}
}

\author{
Estefany Castillo Ventura
\quad
Javier Ulises Solis Lastra
\quad
Anarosa Alves Franco Brandão
\\[0.35em]
\small Escola Politécnica, Universidade de São Paulo (USP)
\\[0.30em]
\small\texttt{\{tecave99,jsolis,anarosa.brandao\}@usp.br}
}

\date{}

\begin{document}

\maketitle


\begin{abstract}
Despite the wide availability of digital resources for teaching mathematics, their effectiveness still depends on selection and integration processes that lack explicit pedagogical criteria. This reveals a gap between their instructional design and their pedagogical use in the classroom. To address this gap, this work advances toward the operationalization of learning theories through observable pedagogical variables. 
The contribution focuses on translating principles derived from learning theories into dimensions and variables structured as metadata, enabling the characterization of digital resources for teaching-support systems. Based on a literature review evidencing the limited level of operationalization of these theories in educational technology, a structure organized into multiple dimensions is proposed. 
This paper addresses one of the multiple dimensions, namely the conceptual structure of the content dimension, showing how aspects such as prior knowledge, representation, and conceptual construction can be represented as metadata. This is intended to support a pedagogically informed selection and integration of resources.
\end{abstract}

\vspace{0.3cm}

\noindent
\textbf{Keywords:}
Instructional design, educational technology, learning technology, P-12, mathematics education.


\section{Introduction}

The increasing availability of digital resources related to mathematics education does not guarantee, by itself, improvements in learning outcomes. While tools such as simulators, dynamic environments, and interactive applications enable new forms of mathematical representation and exploration \cite{41}\cite{110}\cite{104}, their impact depends on how they are selected and integrated into classroom practice, processes that often lack explicit pedagogical criteria \cite{R2}\cite{R4}\cite{R7}.

The literature review conducted in this study shows that learning theories are predominantly used as frameworks for justification, analysis, or design principles, but are rarely operationalized into functional components of technological systems (see Section~\ref{sec:LiteratureReview}). As a consequence, decision-making regarding their effective use falls on teachers, who must select and use educational resources \cite{R8}\cite{R7}\cite{DT1} without explicit mechanisms that link these decisions to theoretical principles.

In response to this gap, this work advances toward a conceptual framework for the pedagogical characterization of digital educational resources in high school's mathematics education. This framework is based on the translation of learning theory principles into observable pedagogical variables structured as metadata, with the aim of supporting more explicit and systematic decision-making.

The framework is organized into four complementary dimensions: (i) conceptual structure of the content; (ii) theoretical–pedagogical alignment; (iii) resource conditions; and (iv) teacher mediation. Given its Work-in-Progress nature, this paper focuses only on the {\it conceptual structure} dimension as a first step toward the operationalization of learning theories in pedagogical practice. The definition of this dimension builds on the analysis of how learning theories are used and operationalized in digital educational technologies.

\section{Methodology}
\label{sec:Method}

The proposed framework is grounded in a structured literature review conducted under a defined protocol, with the aim of analyzing the use and level of operationalization of learning theories in digital educational technologies. This analysis made it possible to identify how learning theories are currently employed and to characterize the absence of intermediate levels of operationalization.

Based on this analysis, learning theories were examined to identify aspects of digital resources that are relevant from a pedagogical perspective. These aspects were then formalized as observable pedagogical variables, defined as properties of the resource that can be explicitly identified in terms of content, representation, and conceptual organization. In this work, the definition of these variables focuses on the conceptual structure of the content dimension.

    \section{Learning Theories}
    \label{sec:LearningTheories}

    Mathematics education has been informed by a range of learning theories that explain how students construct mathematical knowledge and how instruction can support this process \cite{NRC2000}. Among the most influential perspectives are constructivist, sociocultural, and cognitive approaches, as well as models developed specifically within mathematics education.

From a constructivist perspective, learning is understood as an active process in which students reorganize their cognitive structures through interaction with content \cite{Piaget}, while sociocultural approaches emphasize the role of social interaction and mediation in cognitive development \cite{Vygotsky}. Meaningful learning theory highlights the role of prior knowledge in conceptual understanding, stating that new knowledge is incorporated through its relationship with existing cognitive structures \cite{Ausubel}. Within mathematics education, the APOS model (Action--Process--Object--Schema) describes knowledge construction as a progression from actions to more complex conceptual structures \cite{Dubinsky}. In addition, perspectives on semiotic representation emphasize how different forms of representation shape access to mathematical concepts and influence understanding \cite{Duval}.

Together, these theories provide principles for understanding how mathematical knowledge is constructed and how content can be organized to support conceptual understanding, particularly in terms of prior knowledge, representation, and conceptual construction. This, in turn, motivates examining how these aspects are reflected in digital educational resources, particularly in terms of their level of operationalization.

    \section{Literature Review}
    \label{sec:LiteratureReview}

    According to the methodology described previously, a literature review is conducted to analyze how learning theories have been used in high school mathematics education, particularly in the context of digital educational technologies, and to evaluate their level of operationalization in such systems.

\subsection{Review protocol}

The review was guided by two research questions:

\begin{itemize}
\item \textbf{RQ1:} Which learning theories have been applied in the teaching of mathematics in high school education?
\item \textbf{RQ2:} How have these theories been used to underpin digital educational technologies, and what is their level of operationalization?
\end{itemize}

RQ1 served an exploratory role by identifying learning theories in the literature, while the subsequent analysis focused on RQ2, examining their operationalization in the design or use of digital educational technologies.

The search was conducted in Scopus and Web of Science and complemented further with searches in SciELO and the Digital Library of the Brazilian Computer Society (SOL), using terms related to learning theory, mathematics education, secondary education, and digital educational technology.

Peer-reviewed studies were included if they addressed learning theories in high school mathematics education and incorporated digital educational technologies, while studies without explicit theoretical grounding, purely technical works without an educational focus, or those focused exclusively on teacher training were excluded.

After duplicate removal, 276 records were screened by title and abstract. A second filtering stage focused on RQ2 and the level of theoretical operationalization. Thirty-three articles were reviewed in full text, and 14 were retained for final analysis. Additional searches in SciELO and SOL did not identify other studies that meet the inclusion criteria.

\subsubsection{Operationalization Scale}

To analyze the role of theory in the identified studies, a scale of theoretical operationalization was defined. The scale distinguishes different ways in which learning theories influence research, ranging from simple conceptual mention to their implementation as functional components of educational technologies. They are presented in Table \ref{tab:OperationalizationScale}.

\begin{table}[ht]
\centering
\caption{Levels of learning theories operationalization in the analyzed studies}
\label{tab:OperationalizationScale}

\renewcommand{\arraystretch}{1.2}

\begin{tabularx}{\textwidth}{
|>{\centering\arraybackslash}p{1.5cm}
|>{\raggedright\arraybackslash}p{3cm}
|>{\raggedright\arraybackslash}X|
}
\hline

\textbf{Level} &
\textbf{Role} &
\textbf{Description} \\
\hline

L0 &
Mention &
General reference without influence on the study. \\
\hline

L1 &
Justification &
Used to justify the technology or pedagogical approach. \\
\hline

L2 &
Analytical use &
Used to analyze or interpret learning processes. \\
\hline

L3 &
Design principles &
Guides the design of interventions or digital resources. \\
\hline

L4 &
Operational implementation &
Implemented as functional components of the system. \\
\hline

\end{tabularx}
\end{table}

\subsection{Results}
\label{subsec:Results}

In relation to RQ2, the analyzed studies show a clear concentration of learning theories at lower levels of operationalization within educational technologies. Most studies employ learning theories as conceptual justification (L1) or as analytical frameworks to interpret learning processes (L2). In several cases, theories are used to inform the design of digital resources or educational activities (L3). No studies were identified in which learning theories are implemented as functional components within technological systems (L4), as summarized in Table~\ref{tab:theories}.

\begin{table}[ht]
\centering
\caption{Learning theories identified and their level of operationalization}
\label{tab:theories}
\renewcommand{\arraystretch}{1.2}
\begin{tabular}{|l|l|l|}
\hline
\textbf{Theory} & \textbf{Articles} & \textbf{Level} \\
\hline
APOS & \cite{19}, \cite{110} & L2 \\
\hline
Kolb & \cite{30}, \cite{43} & L3 \\
\hline
Cognitive constructivism & \cite{39}, \cite{42}, \cite{151} & L0--L1 \\
\hline
Sociocultural constructivism & \cite{39}, \cite{42}, \cite{239}, \cite{151} & L1, L3 \\
\hline
Self-regulated learning & \cite{39}, \cite{225} & L3 \\
\hline
General constructivism & \cite{41}, \cite{49} & L0--L1 \\
\hline
Bruner's constructivism & \cite{42} & L3 \\
\hline
Embodied cognition & \cite{104} & L3 \\
\hline
Didactic constructivism & \cite{197}, \cite{210} & L1 \\
\hline
\end{tabular}
\end{table}

\subsection{Gaps and Opportunities}
\label{subsec:GapsAndOppotunities}

The absence of studies at the level of operational implementation (L4) evidences a gap between learning theories and their incorporation as functional components within educational technologies. As a consequence, although these theories may influence the design or analysis of digital resources, they are not integrated in forms that enable their direct use in pedagogical decision-making by teachers. This situation reveals the lack of intermediate levels of operationalization that would allow translating theoretical principles into explicit and observable forms in educational practice.

\section{Conceptual Framework}
\label{sec:TheoreticalFramework}
In response to the limited level of operationalization of learning theories identified in the literature, this work advances toward the definition of a conceptual framework aimed at the pedagogical characterization of digital resources in mathematics.

In this sense, the framework translates principles derived from learning theories into observable pedagogical variables. These variables are understood as properties of the resource that can be explicitly identified based on its content, representation, and conceptual organization. In this way, they enable the pedagogical characterization of digital resources and can be represented as metadata.

Through this representation, the framework may provide a basis for the formulation of analysis and selection criteria that support teachers’ decision-making without replacing their professional judgment. 

This work focuses on the conceptual structure of the content as an initial step toward the complete model.

\subsection{Conceptual Structure of the Content}
\label{subsec:ConceptualStructureOfTheContent}

In digital educational resources, the conceptual structure of the content plays a central role, as the way in which it is presented (through different forms of representation and types of interaction) directly shapes the forms of understanding that can be supported \cite{R9}\cite{R8}.

From the perspective of meaningful learning, the incorporation of new concepts depends on their relationship with prior cognitive structures \cite{Ausubel}. This implies identifying the knowledge required to understand the content addressed by the resource, allowing an analysis of its coherence with respect to the learner’s prior knowledge. A resource becomes inadequate when it assumes knowledge that has not been previously acquired, thereby constraining the possibilities for understanding \cite{43}, an aspect captured by the variable {\tt Prior Knowledge Requirements}.

In turn, APOS theory provides a framework for characterizing the type of conceptual construction that a digital resource is designed to support, distinguishing between actions, processes, objects, and schemas \cite{Dubinsky}. Rather than describing internal cognitive states of the learner, this work adopts APOS as an analytical lens to examine the structure of the resource, particulary in terms of interaction, representation, and organization. From this perspective, the focus is on identifying which levels of conceptual construction are made accessible by the design of the resource and by the ways in which it can be used in instructional contexts. \cite{19}\cite{110}. These aspects are reflected by the variable {\tt Type of Conceptual Construction Supported}.

\begin{figure}[ht]
\centering
\includegraphics[width=6cm]{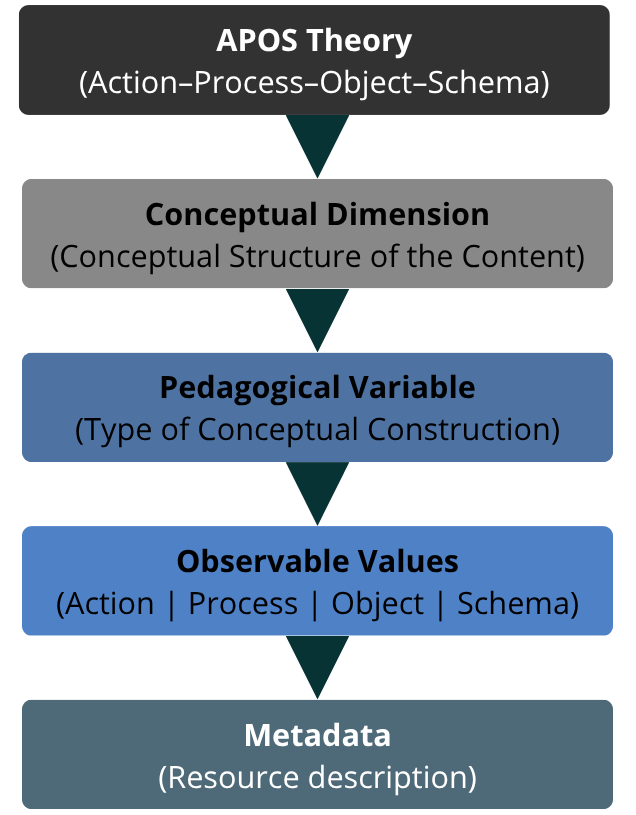}
\caption{From APOS theory to observable pedagogical variables and metadata representation}
\label{fig:APOStheory}
\end{figure}

Figure ~\ref{fig:APOStheory} illustrates how APOS theory informs the definition of observable pedagogical variables, showing how levels of conceptual construction are represented through the structural features of digital resources as metadata.

Likewise, the forms of representation used in a resource play a central role in accessing mathematical concepts. From the perspective of semiotic representation registers, different representations (such as graphical, symbolic, or verbal) provide distinct ways of approaching the same content \cite{Duval}. In this sense, both the predominant representation and the coordination of multiple representations influence the forms of understanding that can be supported \cite{110}.

Taken together, these perspectives support the identification of observable pedagogical variables related to the conceptual structure of the content, namely:

\begin{itemize}
    \item \texttt{Mathematical Content Addressed}
    \item \texttt{Prior Knowledge Requirements}
    \item \texttt{Predominant Representation}
    \item \texttt{Type of Conceptual Construction}
\end{itemize}

\noindent
These variables were derived from the theoretical perspectives discussed in \cite{43}.

Table~\ref{tab:conceptual_structure} presents these variables together with their observable aspects and associated values, establishing a correspondence between learning theories and their representation as metadata.

\begin{table}[ht]
\centering
\caption{Observable pedagogical variables for conceptual content structure}
\label{tab:conceptual_structure}

\renewcommand{\arraystretch}{1.2}

\begin{tabularx}{\textwidth}{
|>{\raggedright\arraybackslash}p{3.5cm}
|>{\raggedright\arraybackslash}X
|>{\raggedright\arraybackslash}X|
}
\hline

\textbf{Variable} &
\textbf{Observable values} &
\textbf{Possible values} \\
\hline

Mathematical Content Addressed &
Topic/subtopic; type of approach &
linear function, derivative, limit; procedural, conceptual, applied \\
\hline

Prior Knowledge Requirements &
Required prior concepts; dependency level &
algebraic operations, notion of variable, graphical interpretation; essential, recommended, not required \\
\hline

Predominant Representation &
Main representation; multiple representations; coordination between representations &
graphical, symbolic, numerical, verbal; yes, no; explicit, implicit, not present \\
\hline

Type of Conceptual Construction &
Predominant level of conceptual construction &
action, process, object, schema \\
\hline

\end{tabularx}
\end{table}


The dimension presented in this section is part of a broader set of dimensions for the pedagogical characterization of digital resources. The {\it conceptual structure of the content} is not sufficient on its own, as its potential to support understanding is closely related to other aspects of resource use in educational contexts \cite{DT1}\cite{R8}.

Within the proposed framework, dimensions such as {\it theoretical–pedagogical alignment}, {\it resource conditions}, and {\it teacher mediation} address complementary aspects of resource use. These include the types of activities supported, the possibilities for interaction and representation, and their integration into instructional sequences \cite{43}\cite{110}\cite{104}\cite{30}. However, given the scope of this work, the analysis focuses on the conceptual dimension as a starting point, leaving the remaining dimensions for future work.

\subsection*{Example: Interactive applet illustrating the secant--tangent transition (\href{https://www.geogebra.org/m/G8VtrntT}{link})}

The resource represents a function graphically together with a secant line defined by two points, one of which varies with a parameter $h$. As $h$ approaches zero, the secant line converges to the tangent line at a given point.

\begin{itemize}
    \item \textbf{Mathematical Content Addressed:} derivative at a point; conceptual.
    \item \textbf{Prior Knowledge Requirements:} notion of function (essential), graphical interpretation (essential), slope of a line (essential); secant and tangent lines (recommended).
    \item \textbf{Predominant Representation:} graphical; multiple representations: yes; coordination: implicit.
    \item \textbf{Type of Conceptual Construction:} process.
\end{itemize}

This resource allows students to explore how the slope of the secant line varies as $h$ changes, supporting the interpretation of the derivative as a limiting process. Although the interaction involves actions (e.g., manipulating the parameter), the main contribution of the resource is making the transition from secant to tangent accessible as a process, consistent with the process level in APOS theory.

\section{Discussion}
\label{sec:discussion}

This study reveals a structural disconnection between learning theories and their effective incorporation into educational technologies. In particular, although these theories are used as frameworks for justification, analysis, or design, they are not operationalized into structures that become part of the internal logic of the systems or that can guide decision-making in pedagogical practice.

As a consequence, aspects related to the pedagogical use of digital resources remain external to the technological artifact and are effectively delegated to the teacher at the moment of use. This creates a misalignment between theoretically grounded design and the absence of explicit structures to support its implementation in practice. 

\subsection{The teacher as the core of pedagogical integration}

In this context, the teacher becomes the central agent in the integration of digital resources, as the interpretation and use of these tools depends on their judgment in practice. \cite{110}\cite{DT1}.

In this scenario, the proposed conceptual framework aims to provide observable pedagogical variables that may serve as a basis for more explicit and systematic analysis of digital resources. In particular, the dimension of conceptual content structure developed in this work supports the analysis of prior requirements, forms of representation, and the type of conceptual construction that a resource may promote. (Section~\ref{subsec:ConceptualStructureOfTheContent})

\subsection{Toward a pedagogy based on explicit criteria}

The proposed framework seeks to make explicit the criteria guiding the pedagogical use of digital resources, articulating them as observable variables for systematic analysis.

It operates as an intermediate layer between theoretical knowledge and teaching practice, translating learning theory principles into explicit forms for resource analysis and selection. In this way, it addresses the identified gap not by embedding theories into systems, but by providing a structured basis that may support their operationalization in practice.

\subsection{Limitations of the study}

This study presents some limitations. The review was restricted to specific databases and criteria, which may have excluded relevant studies. Additionally, the classification of the level of operationalization involves a degree of interpretation.

Finally, the proposed framework constitutes an initial approach. Its usefulness should be evaluated in future research, particularly through its application in real educational contexts and in the development of repositories or recommendation systems for digital resources.

\section{Conclusion}
\label{sec:conclusion}

This work introduces a conceptual framework for the pedagogical characterization of digital resources in mathematics education. A structured literature review revealed a limited operationalization of learning theories in digital technologies, highlighting a gap between pedagogical design and its use in practice.

To address this gap, the framework translates principles from learning theories into observable pedagogical variables, structured as metadata to characterize digital resources. While organized into multiple dimensions (conceptual structure of the content, theoretical–pedagogical alignment, resource conditions, and teacher mediation), this Work in Progress focuses on the conceptual structure dimension as an initial step.

This representation, based on observable pedagogical variables structured as metadata, may support computational processing and enable the organization of digital resources according to these variables. This creates the basis for future work oriented toward the development of tools that support pedagogical decision-making.


\bibliographystyle{plain}
\bibliography{references}

\end{document}